# Giant magnetoelectric effect under high bias fields in cylindrical bilayered Ni-PZT composites

D. A. Pan, J. Lu, Y. Bai, W. Y. Chu and L. J. Qiao*

Environmental Fracture Laboratory of Education of Ministry, Corrosion and Protection Center, University of Science and Technology Beijing, Beijing 100083, P. R. China

*Corresponding author. Fax: +86 10 6233 2345
E-mail address: lqiao@ustb.edu.cn

**Abstract**: For conventional plate and disc layered magnetoelectric (ME) composites, the ME effect appeared under low bias magnetic fields, which caused by line magnetostriction. In this paper, we present a new structure ME composites, cylindrical bilayered composites. Unexpectedly, the ME effect also appeared under high magnetic fields in this cylindrical bilayered Ni-lead zirconate titanate (PZT) composite. We thought this ME effect caused by volume magnetostriction. The results showed that the ME voltage coefficient of the cylindrical bilayered composite of Ni-PZT was the sum of $\alpha_{E(\lambda)}$ caused by line magnetostriction under low bias fields and $\alpha_{E(\omega)}$ induced by volume magnetostriction under high fields. At the resonance frequency, $\alpha_{E,A(\omega)}$ increased linearly with bias magnetic field $H_{DC}$ to 30 V/cm Oe at $H_{DC}$ =8 kOe, and was much larger than $\alpha_{E,A(\lambda)}$ at $H_m$=0.6 kOe. The novel characteristic of ME voltage coefficient increased linearly with the rise of $H_{DC}$ after 1 kOe in axial mode makes the cylindrical composite a promising candidate for application in high magnetic field sensor.

**Keyword:** giant magnetoelectric effect, cylindrical bilayered composite, volume magnetostriction

Multiferroic materials have drawn increasing interest due to their multi-functionality, which provides significant potential for applications in the next-generation multifunctional devices.[1] In the multiferroic materials, the coupling interaction between ferroelectric and ferromagnetic orders could produce some new effects, such as magnetoelectric (ME) or magnetodielectric effect.[2] The ME response, characterized by the appearance of an electric polarization upon applying a magnetic field and/or a magnetization upon applying an electric field, has been observed as an intrinsic effect in some single phase materials.[3,4] Alternatively, multiferroic composites made of ferromagnetics and ferroelectrics were found to exhibit large room-temperature extrinsic ME effects recently,[5-8] which has been known as a product property,[9] i.e. a new property of such composites that either individual component phase does not exhibit. This ME effect can be defined as a coupling of magnetic-mechanical-dielectric behavior. That is, when a magnetic field is applied to the composites, the ferromagnetic phase changes the shape magnetostrictivity, and then the strain is passed to the piezoelectric phase, resulting in an electric polarization.[10]

To achieve better magnetoelectric properties, giant magnetostrictive material $Tb_{1-x}Dy_xFe_{2-y}$ (Terfenol-D) was used to combine with piezoelectric materials, such as lead zirconate titanate (PZT) and polyvinylidene fluoride (PVDF), in a laminate structure.[11-18] The reported ME voltage coefficient of bulk laminate samples were around 5.0 V/cm Oe. To improve the interfacial bonding between the magnetostrictive layer and the piezoelectric layer, we developed ME layered composite by electro-deposition. A high ME voltage coefficient of 16V/cm Oe was obtained in a Ni/PZT/Ni trilayered composite. For these plate or disc samples, ME effect appeared under a low bias magnetic field. Moreover, electro-deposition can be used to fabricate magnetoelectric coupling devices with complex shapes and easily control the thickness of each layer. This method overcomes the limitation of previous preparation method, which only simple shape layers, such as disk, square and rectangle can be made into ME laminates. In this work, the giant ME effects under high bias magnetic fields in a bilayered cylindrical composite of Ni-PZT was studied.

A PZT cylinder with a thickness of 3 mm, inner diameter of 18 mm and outer diameter of 20 mm were polarized at 425 K in an electric field of 30-50 kV/cm along radial direction after electroplating electrodes of Ni. After the inner wall of the cylinder was protected by a silicon rubber, it was bathed in nickel aminosulfonate plating solution to electro-deposition Ni on outer side. After 20 hours of electro-deposition, the thickness of Ni reached about 1 mm as illustrated in Fig. 1. The compositions of the plating solution and processing parameters are listed in Table I. Nickel aminosulfonate plating solution was used because of its advantages such as the solution stability, rapid plating speed and small internal stress.

Both bias magnetic field $H_{DC}$ and an AC field $\delta H$ with a frequency ranging from 1 kHz to 120 kHz were applied along the axis of the cylinder, $\alpha_{E,A}$ is obtained. The voltage $\delta V$ across the wall of the cylinder was amplified and measured via an oscilloscope. The ME voltage coefficient was calculated based on $\alpha_E = \delta V/(t_{PZT} \delta H)$, where $t_{PZT}$ is the thickness of PZT, $\delta H$ is the amplitude of the AC magnetic field generated by Helmholtz coils. In the experiment, $\delta H = 22$ Oe as the amplitude of AC current through the coil is equal to 1A.

The dependence of $\alpha_{E,A}$ on bias magnetic field $H_{DC}$ at $f=1$ kHz for $\delta H$ is shown in Fig. 2. Figure 2 shows that $\alpha_{E,A}$ at 1 kHz has a maximum value at $H_m=0.6$ kOe, and then decreases rapidly, however, increases linearly with increasing $H_{DC}$ after $H_{DC}>3$ kOe.

Afterwards, the frequency dependence of $\alpha_{E,A}$ is measured at the bias field of $H_{DC}=H_m$ and $H_{DC}=6$ kOe, respectively, as shown in Fig. 3. For both $H_{DC}=H_m$ and $H_{DC}=6$ kOe, there are sharp peaks at $f_r=59.9$ kHz. Figure 3 shows that there exist giant ME coupling not only under low bias field but also under high bias field such as 6 kOe. The electromechanical resonance peak under high field of $H_{DC}=6$ kOe is $\alpha_{E,A(6 \text{ kOe})}=21$ V/cm Oe and is much larger than that under low field of $H_m=0.6$ kOe.

The bias field dependence of $\alpha_{E,A}$ at resonance frequency of $f_r=59.9$ kHz is shown in Fig. 4. Figure 4 shows that $\alpha_{E,A}$ at $f_r=59.9$ kHz increases linearly with increasing bias field after $H_{DC}>1$ kOe, and up to 30 V/cm Oe at $H_{DC}=8$ kOe.

For ferromagnetic materials, like Ni, line magnetostriction $\lambda=\Delta l/l$ increases with increasing bias magnetic field of $H_{DC}$, then reaches a saturation value $\lambda_s$ at $H_s$. When $H_{DC} < H_s$, the volume change, i. e. volume magnetostriction $\omega = \Delta V/V$ is too small to be measured, when $H_{DC} > H_s$, however, volume magnetostriction increased with increasing field $H_{DC}$.[19] Under low magnetic field, i. e. $H_{DC} < H_s$, the field dependence of the ME voltage coefficient $\alpha_E$ is determined by the variation of the piezomagnetic coupling $q$ with the field $H_{DC}$,[20] and $\alpha_E$ is proportional to $q$, i. e. to $\delta\lambda/\delta H$, where $\delta\lambda$ is the differential magnetostriction. When $H_{DC}=H_s$, $\lambda=\lambda_s$ and then $\delta\lambda/\delta H = 0$, therefore, the $\alpha_{E(\lambda)}$ caused by the line magnetostriction is equal to zero under high bias field. The volume magnetostriction of ferromagnetic phase Ni under high bias field of $H_{DC} > H_s$ can also generate a strain or stress on the piezoelectric phase of PZT, resulting in increasing the voltage of $\delta V$ across the PZT. Thus, $\alpha_{E(\omega)}$ induced by the volume magnetostriction appears under high bias field, which increases with increasing bias field. The total ME effect is the sum of $\alpha_{E(\lambda)}$ caused by line magnetostriction under low fields and $\alpha_{E(\omega)}$ induced by volume magnetostriction under high fields, i. e. $\alpha_E=\alpha_{E(\lambda)}+\alpha_{E(\omega)}$. For a plate or disc trilayered composites, there is no constraint on the boundary of the ferromagnetic phase, and then no $\alpha_{E(\omega)}$ appears under high field, i. e. $\alpha_E=\alpha_{E(\lambda)}$. For the cylindrical bilayered composite illustrated in Fig. 1, $\alpha_E=\alpha_{E(\lambda)}+\alpha_{E(\omega)}$ and $\alpha_{E(\lambda)}$ plays main role under low bias fields and $\alpha_{E(\omega)}$ under high fields.

As shown in Fig. 4, the linearity of the $\alpha_E$ to the bias magnetic field is 0.99757, in the range of 1-8 kOe. It is greatly remarkable because the linearity decreases significantly when the external magnetic fields are larger than 1 kOe for most conventional high-field magnetic sensors.[21] Due to the limitation of our measurement system we have no data of the $\alpha_E$ under higher bias magnetic fields larger than 8 kOe, but it is believed that the range of high linearity of the cylindrical composite can be wider to a large extend.

In summary, the ME voltage coefficient of a cylindrical bilayered Ni-PZT composite is the sum of $\alpha_{E(\lambda)}$ caused by line magnetostriction plays main under low bias field and $\alpha_{E(\omega)}$ induced by volume magnetostriction plays main under high field. i. e., $\alpha_E=\alpha_{E(\lambda)}+\alpha_{E(\omega)}$. At the resonance frequency, $\alpha_{E,A(\omega)}$ increases linearly with $H_{DC}$ to 30 V/cm Oe at $H_{DC}=8$ kOe and is much larger than $\alpha_{E,A(\lambda)}$ at $H_m=0.6$ kOe. The proportional relationship between $\alpha_{E,A(\omega)}$ and $H_{DC}$ after 1 kOe is hopeful of applying ME composites in the sensor of high magnetic field.

**Acknowledgment**

This work was supported by the program for Changjiang Scholars, Innovative Research Team in University (IRT 0509) and the National Natural Science Foundation of China under Grant No. 50572006.

Table I.  Compositions and process parameters of the nickel electro-deposition

| | |
|---|---|
| Nickel aminosulfonate (g/l) | 600 |
| Nickel chloride (g/l) | 20 |
| Boric acid (g/l) | 20 |
| Sodium lauryl sulfate (g/l) | 0.1 |
| pH | 4 |
| Temperature (°C) | 60 |
| Cathodic current density (A/dm$^2$) | 5 |

**Figure Captions Page**

FIG. 1. Schematic of of the cylindrical bilayered composite of Ni-PZT.

FIG. 2. Bias field $H_{DC}$ dependence of ME effect of $\alpha_{E,A}$ at $f$=1 kHz of $\delta H$ for Ni-PZT cylindrical bilayered composite.

FIG. 3. Frequency dependence of $\alpha_{E,A}$ at $H_m$=0.6 kOe (a) and $H_{DC}$=6 kOe (b) for Ni-PZT cylindrical bilayered composite.

FIG. 4. Bias field $H_{DC}$ dependence of $\alpha_{E,A}$ at the resonance frequency of $f_r$=59.9 kHz for Ni-PZT cylindrical bilayered composite.

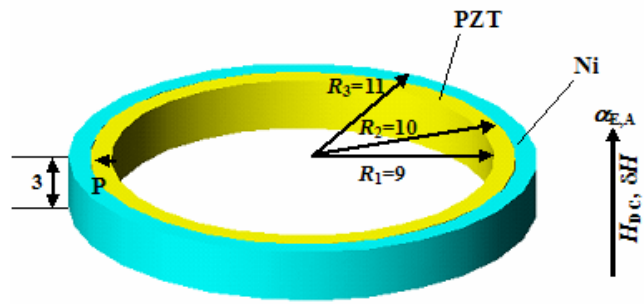

FIG. 1. Schematic of the cylindrical bilayered composite of Ni-PZT.

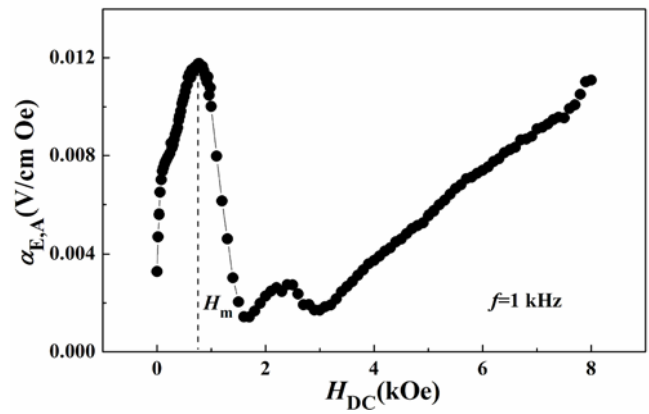

FIG. 2. Bias field $H_{DC}$ dependence of ME effect of $\alpha_{E,A}$ at $f=1$ kHz of $\delta H$ for Ni-PZT cylindrical bilayered composite.

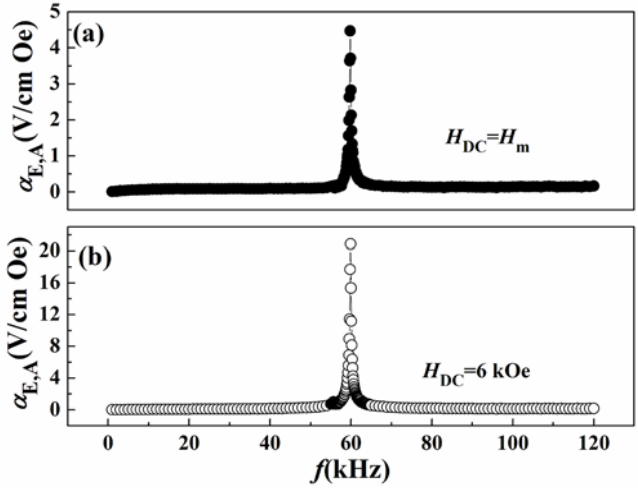

FIG. 3. Frequency dependence of $\alpha_{E,A}$ at $H_m$=0.6 kOe (a) and $H_{DC}$=6 kOe (b) for Ni-PZT cylindrical bilayered composite.

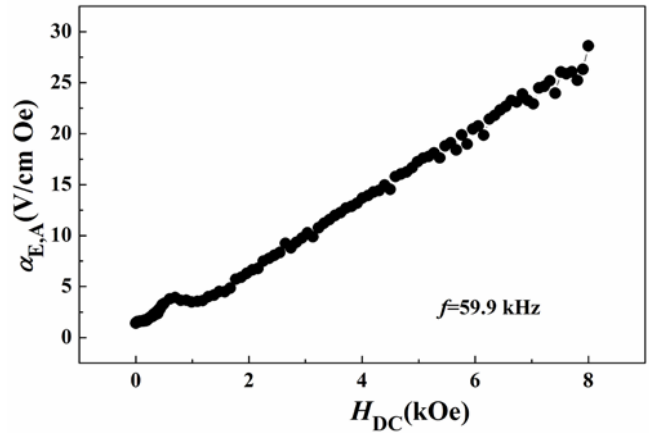

FIG. 4. Bias field $H_{DC}$ dependence of $\alpha_{E,A}$ at the resonance frequency of $f_r$=59.9 kHz for Ni-PZT cylindrical bilayered composite.